\documentclass[twoside,letterpaper]{article}
\usepackage{amsfonts,qic,epsfig}

\begin{document}
\setlength{\textheight}{8.0truein}    

\runninghead{On the Optimal Mean Photon Number for Quantum Cryptography}
            {Pearson and Elliott}

\normalsize\textlineskip
\thispagestyle{empty}
\setcounter{page}{1}

\copyrightheading{0}{0}{2003}{000--000}

\vspace*{0.88truein}

\alphfootnote

\fpage{1}

\centerline{\bf ON THE OPTIMAL MEAN PHOTON NUMBER}
\vspace*{0.035truein}
\centerline{\bf FOR QUANTUM CRYPTOGRAPHY}
\vspace*{0.37truein}
\centerline{\footnotesize DAVID S. PEARSON}
\vspace*{0.015truein}
\centerline{\footnotesize\it BBN Technologies}
\baselineskip=10pt
\centerline{\footnotesize\it Cambridge, Massachusetts 02138, USA}
\baselineskip=10pt
\centerline{\footnotesize\it dpearson@bbn.com}
\vspace*{10pt}
\centerline{\footnotesize CHIP ELLIOTT}
\vspace*{0.015truein}
\centerline{\footnotesize\it BBN Technologies}
\baselineskip=10pt
\centerline{\footnotesize\it Cambridge, Massachusetts 02138, USA}
\baselineskip=10pt
\centerline{\footnotesize\it celliott@bbn.com}
\vspace*{0.225truein}
\publisher{(received date)}{(revised date)}

\vspace*{0.21truein}

\abstracts{
The optimal mean photon number ($\mu$) for quantum cryptography
is the average photon number per transmitted pulse that results
in the highest delivery rate of distilled cryptographic key bits,
given a specific system scenario and set of assumptions about
Eve's capabilities. Although many experimental systems have employed
a mean photon number ($\mu$) of 0.1 in practice, several research
teams have pointed out that this value is somewhat arbitrary.
In fact, various optimal values for $\mu$ have been described in
the literature.
}{
In this paper we offer a detailed analytic model for an experimental,
fiber-based quantum cryptographic system, and an explicit set
of reasonable assumptions about Eve's current technical capabilities.
We explicitly model total system behavior ranging from physical
effects to the results of quantum cryptographic protocols such
as error correction and privacy amplification. We then derive
the optimal photon number ($\mu$) for this system in a range of
scenarios. One interesting result is that $\mu \approx 1.1$ is optimal
for a wide range of realistic, fiber-based QKD systems; in fact,
it provides nearly 10 times the distilled throughput of systems
that employ a more conventional $\mu = 0.1$, without any adverse
affect on system security, as judged against a set of reasonable
assumptions about Eve's current capabilities.
}{}

\vspace*{10pt}

\keywords{Quantum cryptography, Quantum key distribution, Mean photon number}
\vspace*{3pt}
\communicate{to be filled by the Editorial}

\vspace*{1pt}\textlineskip    

\section{Background and Problem Statement}

   It now seems likely that Quantum Key Distribution (QKD)
techniques can provide practical building blocks for highly secure
networks, and in fact may offer valuable cryptographic services, such
as unbounded secrecy lifetimes, that can be difficult to achieve by
other techniques. Accordingly, a number of commercial and research
organizations have begun to build and operate complete QKD
systems. As quantum cryptography has started the transition from
laboratory demonstrations to working systems in the field, questions
of operating efficiency and realistic levels of security have taken on a
heightened importance.\footnote{The opinions expressed in this article
are those of the authors alone,
and do not necessarily reflect the views of the United States
Department of Defense, DARPA, or the United States Air Force.}

   A wide range of techniques has been proposed for quantum
cryptography, and many have been experimentally demonstrated; see
Gisin et al.~\cite{Gisin02} for a superb overview.
However, in realistic settings,
such as operation through the atmosphere or through tens of
kilometers of telecommunications fiber, even the most efficient of
these techniques currently provide no more than roughly 1,000
distilled\footnote{In our terminology, a ``distilled''
key has been sifted, error corrected, and privacy
amplified, and is thus ready to use as key material.}
bits per second, depending on channel losses, choice of
eavesdropping threat model, and a large number of technical
parameters. While this key generation rate is more than sufficient for
rapid rekeying of conventional cryptographic algorithms---for
example it allows rekeying of an AES algorithm with fresh 256-bit
keys 4 times per second---a faster key generation rate would allow a
large number of cryptographically protected traffic flows to be
rekeyed at a given rate. In addition, it is far too low for most uses of
one-time pads. An important practical question, therefore, is how to
increase this rate.

   A number of different approaches may contribute to improved
distilled key generation rates: detector efficiencies may be improved,
e.g., by novel forms of detectors; pulse rates may be increased; and
entropy estimates may be refined so that less privacy amplification is
required for a given observed level of noise. Promising efforts are
underway in all these areas (e.g.~\cite{SSPD01, NIST02, Myers04}).
This paper explores yet
another avenue to increasing the key generation rate, namely, by
finding an optimal value for the mean number of photons emitted in
each pulse, i.e., that which maximizes the distilled key generation rate
for a given scenario and set of eavesdropping assumptions. This mean
photon number is often designated $\mu$ in the QKD literature.

   This paper provides a detailed, quantitative analysis of the
interaction between $\mu$, channel attenuation, and privacy-amplified key
generation rates, and compares the results with prior research on
optimal mean photon number. We specifically consider a phase-modulated
system, with attenuated laser source and cooled InGaAs
APDs, designed for telecommunications fiber; however the results
can be readily generalized to other systems.

   One interesting result is that $\mu \approx 1.1$ is optimal for a
wide range of realistic, fiber-based QKD systems under a reasonable
eavesdropping threat model; in fact, it provides nearly 10 times the
distilled throughput of systems that employ a more conventional $\mu=0.1$,
without any adverse affect on system security. For many steeped
in the field, it may seem counter-intuitive---even downright
false---that a mean
photon number as large as 1, let alone greater than 1, may
be possible without sacrificing all security. However, as Prof. Gisin et
al. have noted in their magisterial survey of quantum cryptography
\cite{Gisin02}, ``multiphoton pulses do not
necessarily constitute a threat to key
security, but they limit the key creation rate because they imply that
more bits must be discarded during key distillation.'' This paper may
be viewed as an elaboration, and preliminary quantification, of that
important remark.

\section{Review of the Current Art}

Although most practitioners of quantum cryptography have now
converged upon a mean photon number ($\mu$) of 0.1 as a good
benchmark value, ``contrary to a frequent misconception, there is
nothing special about a $\mu$ value of 0.1, even though it has been
selected by most experimentalists. The optimal value---i.e., the value
that yields the highest key exchange rate after distillation---depends
on the optical losses in the channel and on assumptions about Eve's
technology.''~\cite{Gisin02}
In fact, in recent years, at least three leading research
teams have carefully investigated the optimal mean photon rate, and
have come to differing conclusions. Accordingly, this section
recapitulates both the widespread rationale for $\mu=0.1$, and the
previous research on the relationship of $\mu$ to distilled key rate.

   As will be seen, for some years the QKD community has held, in
effect, an ongoing discussion of the optimal mean photon number for
various contexts, but generally as side comments within papers
devoted to other topics. As a result, it has been difficult to find a
detailed and explicit linkage between eavesdropping threat models
and optimal mean photon numbers.

   The origin of the value 0.1 for the mean photon number was the
very first experimental realization of QKD by Bennett et al. in 1992
\cite{BBBSS92}. This early work analyzed various kinds of attacks on the small
number of multi-photon pulses produced, including one version of
unambiguous state discrimination, and concluded that unambiguous
state discrimination was impossible for such a small $\mu$ without
significantly biasing the detector statistics at Bob. Later researchers
have shown that this conclusion is incorrect~\cite{GH00}. However, the
number of bits that Eve can discover is very small, and Bennett et al.
left a significant safety margin in their estimate. The attacks they
considered feasible involved intercepting one photon from each
multi-photon pulse and measuring it.  For each such pulse that
reaches Bob, they assume Eve gains one bit of information, thus
implicitly allowing Eve to have a quantum memory and to measure
the photon only when its basis is disclosed.

   Many other experimental systems, including ours, borrowed from
\cite{BBBSS92} the value of 0.1 for $\mu$ as well
as the estimate of Eve's advantage
from photon-number splitting (PNS) attacks. Two experimental
teams, from Los Alamos~\cite{LANL00} and
IBM Almaden~\cite{Bethune04}, then calculated
optimal numerical values for $\mu$ in their systems, based on this
estimate.  For the free-space system used by Los Alamos, this value
of $\mu$ was 0.4, while for IBM's ``plug-and-play'' fiber-based system it
was 0.3. The IBM Almaden group also examined the throughput vs.
mean photon number for a number of different eavesdropping
models.

   On the theoretical side, Gilbert and Hamrick~\cite{GH00} performed an
extensive analysis of possible attacks on multi-photon pulses,
including splitting, unambiguous state discrimination, and
surreptitiously replacing the channel to Bob with a perfectly
transparent one.  In short, they selected a more formidable
eavesdropping model than posited in the analyses of Bennett, Los
Alamos, or IBM Almaden. Granting Eve such powers, they produced
a much more conservative estimate of the amount of information Eve
might gain. They also analyzed the optimum mean photon number in
one specific scenario, an aircraft to a LEO satellite, and found it to be
0.455, although in this case they allow Eve less power than in a fiber
link---specifically, she is not able to replace the channel with a
lossless one.

   These differing estimates are further compared in section 5.

\section{Our Analytical Approach}

Our analysis, in the following section, is derived from a moderately
detailed mathematical model of a full QKD system for use in
telecommunications fiber, including both physical effects and the
outcomes of higher level protocols, validated against two working
systems in the laboratory. This section briefly describes our working
systems (the concrete subjects of analysis), then discusses the major
elements in our analytic model. Appendix A contains the full text of
the model.

\subsection{Functioning Systems for Quantum Cryptography}

   BBN, Boston University, and Harvard University are currently
building a large-scale quantum cryptography system, the DARPA
Quantum Network, and fielding it into dark fiber in the Cambridge,
Massachusetts metropolitan area.
See for example~\cite{Elliott02, BBN03} for details
on this network and its design goals. Two interoperable QKD systems
in the DARPA Quantum Network started 24x7 duty in October 2003;
we call these 'Mark 2' systems because they replaced our
first-generation link, which started continuous operation in December
2002. These systems were inspired by a pioneering Los Alamos
system~\cite{LANL96} and designed to run through telecommunications fiber as
widely deployed today.

   Each Mark 2 system employs a highly attenuated
telecommunications laser at 1550.12 nm, phase modulation via
unbalanced Mach-Zehnder interferometers, and thermo-electrically
cooled InGaAs avalanche photo detectors (APDs). Most Mark 2
electronics are implemented by discrete components such as pulse
generators. At present, incoming dim pulses are detected by Epitaxx
EPM 239 AA APDs cooled to approximately --40 degrees Centigrade
and gated during a pulse arrival period. Since custom cooling and
electronics are required, we designed and built our own cooler
package to maintain the APDs at the requisite operating temperatures.
Even with this special treatment, they suffer considerably from low
Quantum Efficiency (QE), relatively high dark noise, and serious
after-pulsing problems. These cooled detectors form one of the most
important bottlenecks in the overall system performance, as they
require on the order of 10 $\mu$sec to recover between detection events.
The overall link has been designed to run at up to 5 Mb/s transmit
rate but with a dead-time circuit to disable the APD after a detection
event in order to accommodate this recovery interval and suppress
detector after-pulsing.

   BBN's QKD protocol stack is an industrial-strength
implementation written in the C programming language for ready
portability to embedded real-time systems. At present all protocol
control messages are conveyed in IP datagrams so that control traffic
can be conveyed via an internet. Two aspects of BBN's QKD
protocol stack deserve special mention. First, it implements a
complete suite of QKD protocols. In fact, it implements multiple
``plug compatible'' versions of some functions, e.g., it provides both
the traditional BB84 sifting protocol and the newer ``Geneva'' style
sifting~\cite{SARG03}.
It also provides a choice of entropy estimation functions
including the well-known BBBSS92 estimates~\cite{BBBSS92}, Slutsky's defense
frontier analysis~\cite{Slut98}, and the newer
Myers-Pearson estimate~\cite{Myers04}. We
expect to add additional options and variants as they are developed.
Second, BBN's QKD protocols have been carefully designed to make
it as easy as possible to plug in other QKD systems, i.e., to facilitate
the introduction of QKD links from other research teams into the
overall DARPA Quantum Network.

\subsection{Analytic Tools used in this Paper}

Over the past two years, we have developed a Matlab / Octave model
to analyze the expected efficiency of current and projected
fiber-based QKD systems in the DARPA Quantum Network. The complete
model is provided in Appendix A. Some aspects of the model have
been derived from the QKD literature, but most have been developed
from first principles. Dr. John Myers of Harvard University has
provided many of the equations in this model; the authors have
provided the remainder. Of course the authors are solely responsible
for any flaws in this published model.

   This model provides for a wide range of input parameters such as
pulse rate, mean photon number at Alice, attenuation, detector
efficiency, dark count, and after-pulsing characteristics, residual
phase error in the Mach-Zehnder interferometers, and so forth. It also
provides input parameters for higher layers of the QKD protocol
stack, such as the sifting protocol employed, information revealed
during error detection and correction, entropy estimation technique,
etc. We briefly discuss these inputs, and the associated calculations,
in the following paragraphs. Although the model provides basic
estimates for a range of physical and protocol phenomena, it is by no
means complete. For example, it does not include any
characterization of stray light, of chromatic or polarization mode
dispersion, and so forth. However, the current version of this model
has been validated against our QKD systems running both through a
fiber spool in the laboratory and through a 17km fiber loop between
BBN and Harvard University, and its results agree well with
experimental measurement. Thus it appears to capture at least the
most important drivers for realistic system behavior.

   As shown in Appendix A, the model inputs represent a fiber-based
system with a 5 Mb/s pulse rate, 0.1 mean photon number ($\mu$),
operating through 10.55 km of telecommunications fiber with an
overall fiber attenuation of 2.5 dB. The average receiver loss factor is
10.4 dB, with a residual phase error in the Mach-Zehnder
interferometers of 3 degrees after both passive and active path length
stabilization. The path length stabilization and framing overhead
results in a duty cycle of 80\% for usable QKD bits. Detector
efficiency is 13\%, with mis-steered light occurring in 0.9\% of the
detections, and a dark count probability of $2.8\times 10^{-5}$ per pulse. At
higher layers of the QKD protocol stack, the traditional BB84 sifting
algorithm is modeled, with the BBN variant of the Cascade error
detection and correction protocol using a block size of 4,096 bits with
64 sets, the traditional BBBSS92~\cite{BBBSS92} entropy
estimate, and a residual
confidence level (the probability that Eve has more information than
estimated) of $10^{-6}$. These values capture the current state of our QKD
systems as of January 2004.

   It should be apparent from inspection of Appendix A that these
parameters can be readily adjusted to model other fiber-based
systems, e.g., different detector characteristics, protocol behavior, and
so forth. One could also extend the model to free-space systems, or
systems based on pairs of entangled photons, but this would require
that additional equations be added to the model rather than mere
adjustment of input parameters.

\section{Eavesdropping Model and Defense Function}

   The most critical factor driving an optimal choice of mean
photon number is determining what sort of attacks Eve can employ.
For intercept-resend attacks on the single-photon pulses, there is a
fairly well-developed theory about how much privacy amplification is
necessary~\cite{Slut98, Myers04}. For multi-photon pulses, a number
of possible attacks have been proposed and
analyzed~\cite{GH00, BLMS00, LutPRA}, but it is by no
means clear that the list of possible attacks is complete yet~\cite{Curty}.
Many of the theoretically possible attacks are very far from practical
implementation with current technology.

    Note that these assumptions about Eve's abilities must be built
into the privacy amplification margin used in any working QKD
system, so they are by no means idle questions. If one wishes to
deploy QKD securely, one must choose these assumptions carefully.
Once we have chosen these assumptions and the privacy
amplification formula, numerical optimization techniques can
determine the optimal multi-photon probability. Therefore it is useful
to explicitly list a set of assumed capabilities for Eve for a given
scenario, as the rates vary greatly depending on the assumptions.

    We must decide, for example, whether we wish to guard against
an Eve possessing the capabilities listed in Table~\ref{tab:threat}.
Many research
results assume that Eve possesses all these capabilities; for some
papers it is difficult to determine exactly which capabilities are
assumed.

\begin{table}[htb]
\begin{center}
\tcaption{Eavesdropping model used in this analysis.}\label{tab:threat}
\vspace{.5em}
\renewcommand{\arraystretch}{2}
\footnotesize
\begin{tabular}{|c|p{3.5in}|}
\hline
Eve Has? & Potential Technological Capabilities for Eve \\ \hline
\Large\checkmark & Perfect detectors \\ \hline
\Large\checkmark & A perfect long-term quantum memory \\ \hline
\Large\checkmark & Adaptive beam-splitters, which split
at most one photon from the signal~\cite{Curty} \\ \hline
\Large\checkmark & Reliable quantum non-demolition
measurement of the total number of photons \\ \hline
& The ability to perform unambiguous state discrimination
on pulses with 3 or more photons \\ \hline
& The ability to discriminate multi-photon pulses in
intercept/resend attacks~\cite{Gisin02} \\ \hline
& The ability to substitute low or zero-loss fiber, or to
perform quantum teleportation with small loss \\ \hline
\end{tabular}
\end{center}
\end{table}

   It is our belief, following Gisin, et al.~\cite{Gisin02},
that it is reasonable to
guard against eavesdropping that is currently feasible, or may be in
the not-too-distant future, rather than make deployment infeasible by
attempting to guard against theoretical attacks that may never be
possible. Note, in particular, that near-perfect detectors, particularly if
they can resolve the number of photons in a pulse, adaptive
beam-splitters, or quantum non-demolition (QND) measurements can all
give us a reliable way to build a true single-photon source, which
would, in turn, render PNS attacks harmless. QKD is very likely to
shift to true single-photon emitters long before we need to worry
about an eavesdropper with a long-term quantum memory.  It is one
of the greatest virtues of QKD that, unlike classical cryptography,
there is no risk that a future powerful adversary endangers our
communications in the present.

   Accordingly, the check marks in Table~\ref{tab:threat} indicate which
technology we assume Eve has for the purposes of this analysis, and
for the current operation of our working QKD systems. We believe
that these assumptions are reasonable for current scenarios, since
many of the postulated technologies appear to be beyond today's
current state of the art.

   Finally, given this explicit set of assumptions about Eve's current
capabilities, one must select an entropy estimate used as input for
privacy amplification. This entropy estimate includes Eve's
information from intercept-resend attacks, called by Slutsky et al. the
``defense function''~\cite{Slut98}. Here we use results based on the original
entropy estimate in BBBSS92, but our analytic model explicitly
calculates three different entropy estimates (BBBSS92~\cite{BBBSS92},
Slutsky \cite{Slut98}, Myers-Pearson~\cite{Myers04}). The choice of
optimal mean photon number is very similar for all choices of entropy
estimate.

\section{Results and Discussions}

   Given all these assumptions, we can employ an analytic model
(Appendix A) to calculate the optimal mean photon number ($\mu$) over
a range of scenarios.  Recall that the ``optimal'' value is that which
maximizes the delivery rate of distilled bits / second, i.e., optimizes
across the system-wide effects of multi-photon emission probabilities,
attenuation, dark noise, sifting, bits revealed during error detection
and correction, and the necessary amount of privacy amplification.

\begin{figure}[tbh]\begin{center}
\epsfig{file=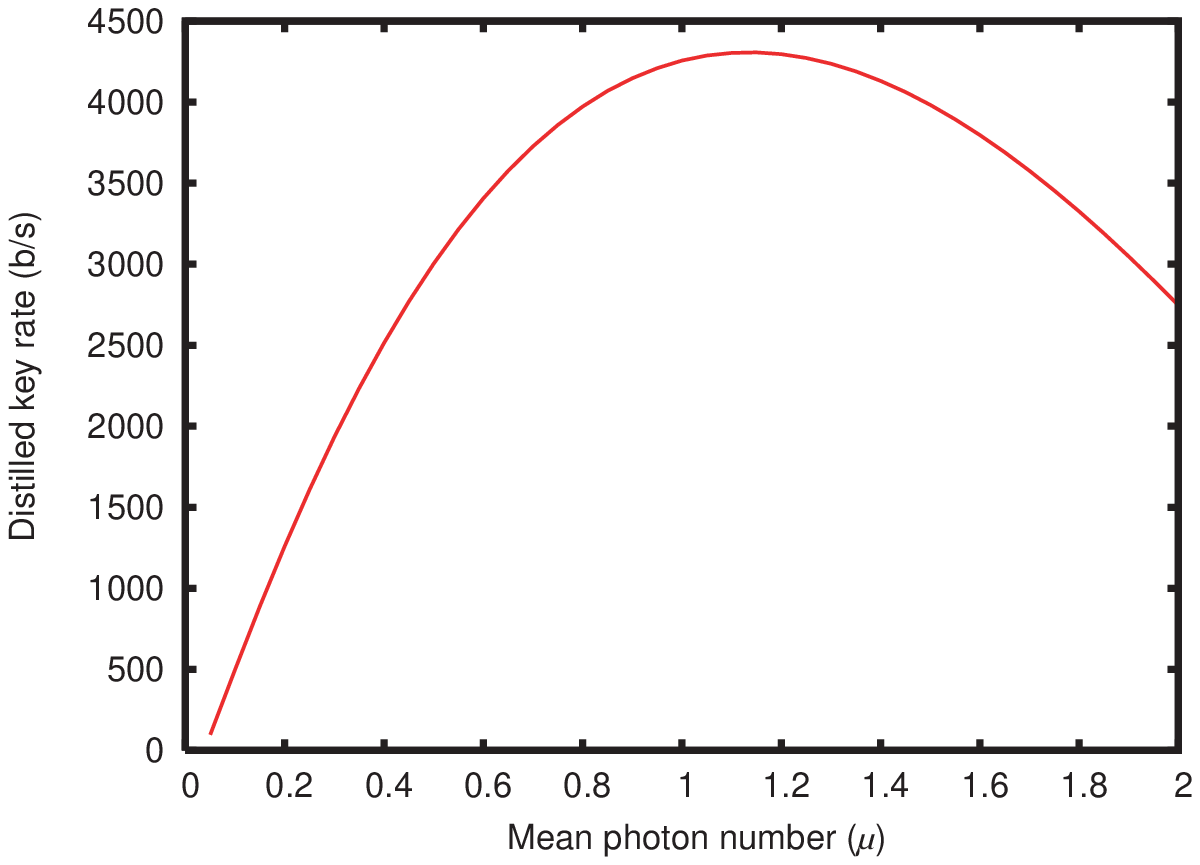, width=4.75in}
\vspace*{8pt}
\fcaption{Distilled Key Rate as a Function of Mean Photon
Number ($\mu$) for a 10.55 km fiber link with 2.5 dB loss.}
\label{fig:tmu}
\end{center}\end{figure}

   The model allows us to extrapolate system performance in a
number of scenarios, e.g. if we had longer fibers, a faster pulse rate,
or better detectors. In particular, we can analyze the effects of
changing the mean photon number. In Figure~\ref{fig:tmu} we vary only the
mean photon number $\mu$, with all other parameters derived from one of
our current QKD systems (with 10.55 km of optical fiber between
Alice and Bob). It is very apparent that the current mean photon
number $\mu$, approximately 0.1 photon, is far from optimal in this
setting. Instead the mean photon number $\mu$ should be slightly more
than 1 (about 1.15) to achieve the optimal distilled key rate.

\begin{figure}[tbh]\begin{center}
\epsfig{file=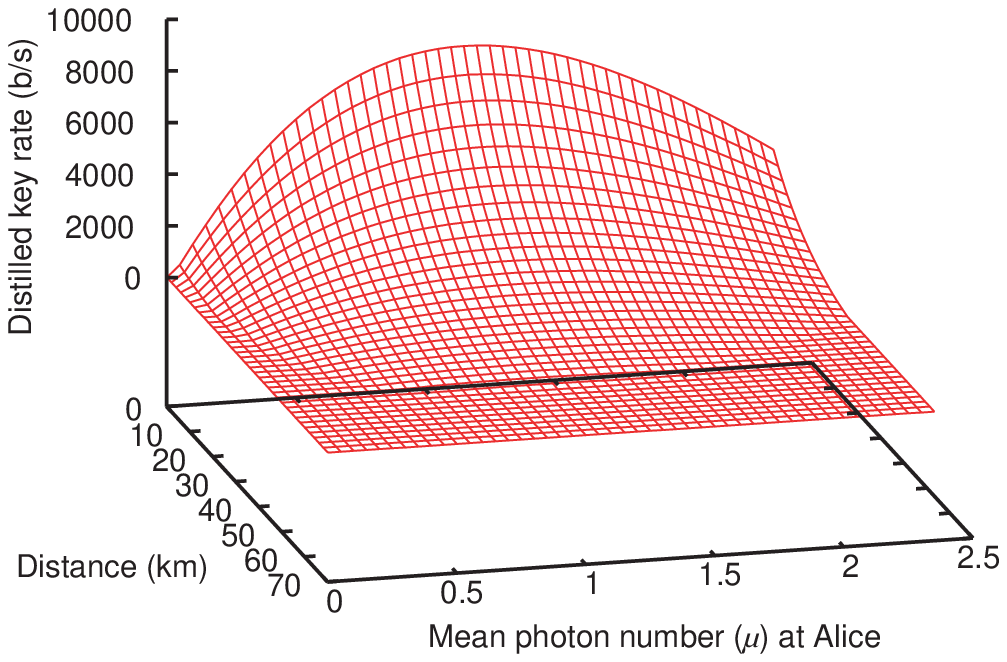, width=4in}
\vspace*{8pt}
\fcaption{Distilled Key Rate as a Function of Distance and
Mean Photon Number ($\mu$).}
\label{fig:surf}
\end{center}\end{figure}

   Another major objective in optimizing $\mu$ is to maximize the
distance available for practical QKD over metropolitan fiber. Figure~\ref{fig:surf}
shows how the distilled key rate varies with both fiber length and $\mu$,
again given specific system characteristics (Appendix A) and the
eavesdropping model of Section 4.

   As can be seen, the distilled key rate falls off dramatically with
distance, and requires high values of $\mu$ for long distances. These
specific results are driven by the relatively low quantum efficiency,
and relatively high dark count, of our current InGaAs detector suite,
but the phenomenon is more general. Larger $\mu$ naturally leads to more
photons at the receiver, and correspondingly more raw key bits per
second, but more importantly it keeps the valid detect rate high
compared to receiver dark noise. Dark noise with a highly attenuating
channel decreases the distilled rate in a very dramatic way because it
translates directly into a higher error rate. The error detection and
correction protocol, such as Cascade, then must reveal a substantial
amount of information to correct the errors. Since it must be assumed,
conservatively, that all these errors are due to eavesdropping, the
estimate of the remaining entropy in the bits drops sharply.

\begin{figure}[tbh]\begin{center}
\epsfig{file=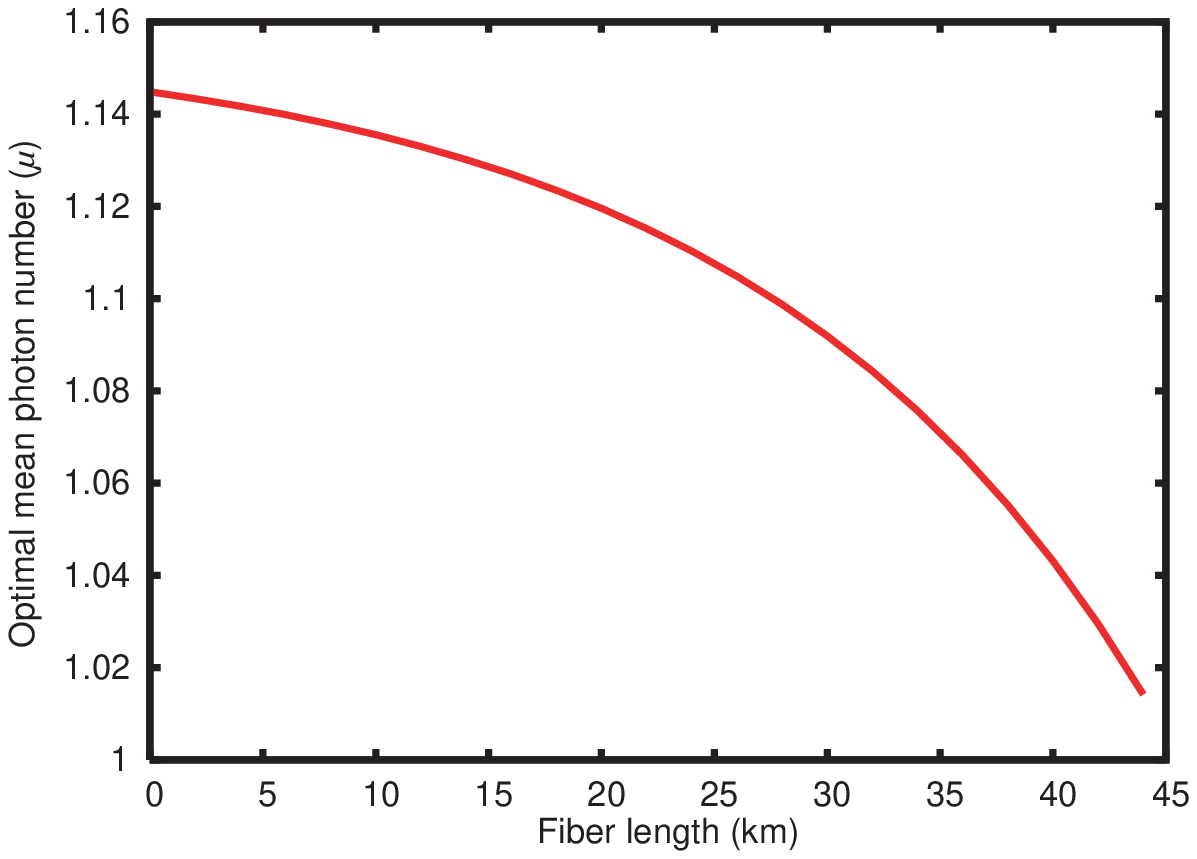, width=4.75in}
\vspace*{8pt}
\fcaption{Optimal Mean Photon Number ($\mu$) as a Function of Fiber Length.}
\label{fig:dmu}
\end{center}\end{figure}

   Since many factors affect the distilled key rate, it is not surprising
that there is not a single optimum value of $\mu$ to employ in all
scenarios. However, for our systems, the optimum value does not
vary by much. Figure~\ref{fig:dmu} shows the optimum $\mu$ for distances from zero
to 50 km. The optimum varies by less than 20\%, from about 1 to 1.2.
The peak of the key rate curve (Figure~\ref{fig:tmu}) is rather broad, so choosing
a value of 1.0, say, for $\mu$ seems to be applicable for a broad range of
operating conditions.

   Since our estimates of the optimal mean photon number are quite
different from conventional wisdom, careful review of the
assumptions and calculations is in order. We believe that Bennett,
Los Alamos, IBM Almaden, and this paper all employ similar
eavesdropping models. This is important, because the eavesdropping
threat model drives the calculations of optimal mean photon number.

\begin{figure}[tbh]\begin{center}
\epsfig{file=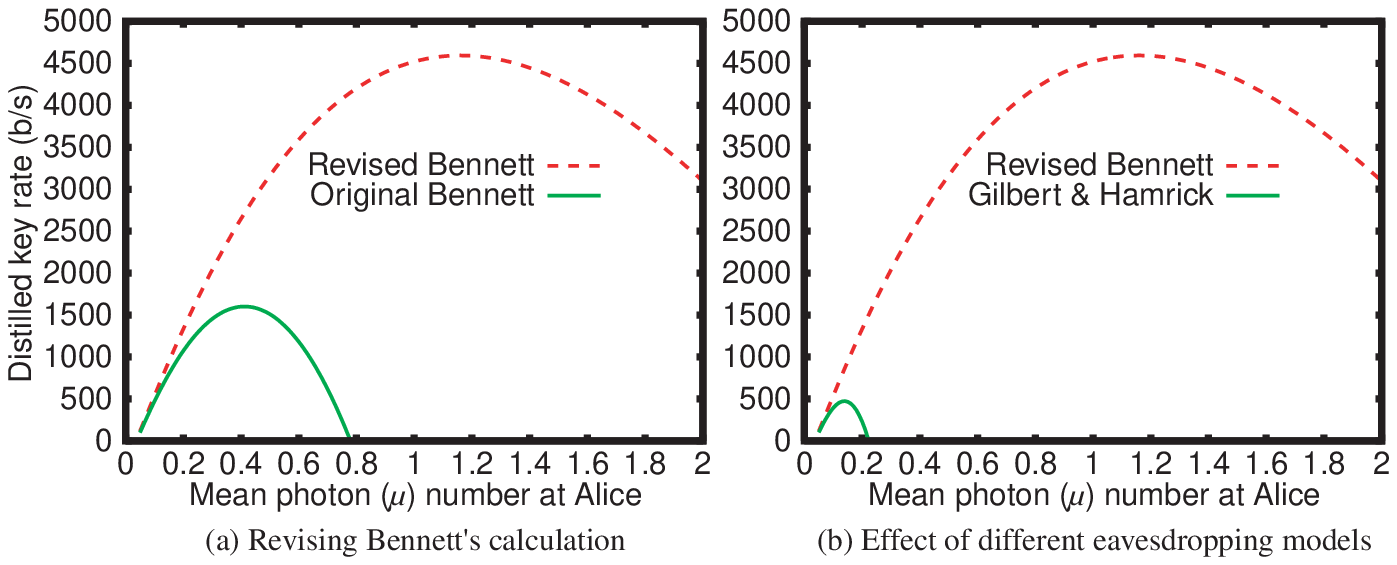, width=400pt}
\vspace*{8pt}
\fcaption{The effect of different eavesdropping estimates on the
distilled key rate as a function of $\mu$ for a 10.55 km fiber link
with 2.5 dB loss.}\label{fig:comp}
\end{center}\end{figure}

   The main difference in our calculation from those of Los Alamos
is as follows. They used, following Bennett et al., a fairly rough
estimate for the fraction of detected pulses that are vulnerable to
splitting. Bennett's (intentionally conservative) estimate was that Eve
could learn a fraction $\mu$ of the bits through beamsplitting.  This
obviously would never allow $\mu=1$, since then Eve would learn all of
the bits.  But this is quite conservative indeed, since the fraction m of
non-empty pulses that contain multiple photons (all of which we want
to assume Eve intercepts) may be more precisely estimated by the
Poisson distribution,
\begin{equation}
m=1-\frac{\mu e^{-\mu}}{1-e^{-\mu}}
\end{equation}
This fraction m is close to $\mu/2$ when $\mu$ is small, but diverges
farther from $\mu$ at higher values.  Figure~\ref{fig:comp}(a)
shows the effect of this
difference between the estimates on distilled key rate for the specific
scenario depicted in Figure~\ref{fig:tmu}. The estimate we use
throughout this paper
(``revised Bennett'') is $mN+\sqrt{2}\,\mbox{erf}\,^{-1}(c)\sqrt{Nm(1-m)}$
where $m$ is defined above, and $c=10^{-6}$ is a confidence parameter,
the residual probability that Eve might gain more information from
multi-photon pulses. The original BBBSS92~\cite{BBBSS92} estimate (``original
Bennett'') is identical except for using $m=\mu$.We are not the first to
employ this revised estimate. Both L\"{u}tkenhaus~\cite{Lut00} and Gilbert and
Hamrick~\cite{GH00} derive their results with the correct multiphoton Poisson
statistics, and indeed predict that for low loss and high efficiency
detectors, the optimum efficiency is achieved for mean photon
numbers greater than 1. Without much discussion, the IBM Almaden
results~\cite{Bethune04}
included curves for the ``revised Bennett'' estimate as well as the
``original Bennett'' estimate for a range of detector efficiencies and
channel losses, in a most interesting graph of the effect of $\mu$ on
distilled key rate in other eavesdropping models, including those of
\cite{GH00} and~\cite{Lut00}. These graphs showed that under some
circumstances $\mu > 1$ is optimal in these other models.

   Figure~\ref{fig:comp}(b) shows the effect of using Gilbert and Hamrick's
estimate~\cite{GH00}, based on a more severe eavesdropping model.
Since they allow Eve perfect unambiguous state
discrimination attacks and zero-loss fiber, it is not surprising that their
estimate results in a far lower key rate.

   The threat model treated in this paper has been implicitly
assumed in the eavesdropping estimates for multi-photon pulses
provided by other research teams~\cite{BBBSS92, LANL00, Bethune04}.
We believe it is a plausible
threat model, given current technology. It is, however, important to
realize that with larger values of $\mu$ we are moving out of the ``comfort
zone'' of these assumptions.  Certain attacks that aren't readily
feasible at small $\mu$ become easier at $\mu=1$. For example,
Bennett et al.  considered a special case of unambiguous
state discrimination in~\cite{BBBSS92},
splitting incoming pulses and measuring one portion in each basis. In
some cases of 3 or more photon pulses, the measurement would result
in both detectors firing in one basis and one firing in the other. When
this happens, Eve can generate a new signal (close to Bob) without
introducing any errors. For small values of $\mu$, Bennett et al. concluded
this attack was harmless. However, when $\mu=1$ and with perfect
detectors for Eve, this attack becomes feasible with a fiber loss of
about 18 dB, corresponding to approximately 90km of fiber at 0.2
dB/km attenuation.\footnote{For this analysis, we assume,
following Gilbert and Hamrick~\cite{GH00}, that Bob is watching for
anomalously high numbers of double detections (when both
detectors click). Without this precaution, Eve would be able
to send more than single-photon signals to Bob after successfully
determining the state, and the attack would be feasible if
the total attenuation, including Bob's receiver, was 18dB.}
Another attack examined by Gisin et al.~\cite{Gisin02}
involves improving the odds of intercept/resend attacks by splitting
the beam, measuring each half in a different basis, and using detectors
that can determine the number of photons in the signal.  In certain
operating regimes (small $\mu$ or short fiber length) this attack is no
better than traditional intercept/resend, and we may use the same
defense function. However by changing the defense function
appropriately (i.e. granting Eve more information for each error bit
received), one can in fact operate safely with a larger mean photon
number. For the operating configuration analyzed in this paper, the
result is still an optimal value of $\mu \approx 1.1$.

\section{Conclusions and Future Work}

In this paper we offer a detailed analytic model for an experimental,
fiber-based quantum cryptographic system, and a set of reasonable
assumptions about Eve's current technical capabilities. We explicitly
model total system behavior ranging from physical effects to the
results of quantum cryptographic protocols such as error correction
and privacy amplification. We then derive the optimal photon number
($\mu$) for this system in a range of scenarios. One interesting result is
that $\mu \approx 1.1$ is optimal for a wide range of realistic, fiber-based
QKD systems; in fact, it provides about 10 times the distilled throughput of
systems that employ a more conventional $\mu=0.1$, without any
adverse affect on system security, given an explicit set of reasonable
assumptions about Eve's current capabilities.

    This paper takes one more step in the ongoing exploration of
optimal mean photon number for a realistic system. Looking ahead,
careful specification of a whole range of eavesdropping threats, and
necessary countermeasures, and of the quantitative effects of each
potential threat model, will be required before QKD can be trusted in
practice. Broadly accepted analysis of a wider range of eavesdropping
techniques, under a range of technologies available to Alice, Bob, and
Eve, is thus desirable.

\nonumsection{Acknowledgements}

We would like to thank
Dr.  Gerry Gilbert and Dr. Mike Hamrick for detailed discussions of
our models and results, and
Dr. Charles Bennett, Dr. Donald Bethune,
and Dr. Richard Hughes for their
very helpful comments on an earlier draft of this paper.
Their prompt
and judicious corrections improved it considerably. Any remaining
errors are, of course, the responsibility of the authors. We also thank
Dr. Martin Jaspan of Boston University for his suggestions on how to
improve our detector analysis.

We are deeply indebted to Dr. Mike Foster (DARPA IPTO) and Dr.
Don Nicholson (Air Force Research Laboratory) who are the sponsor
and agent, respectively, for this research project. This paper reflects
highly collaborative work between the project members. Of these,
particular credit is due to Profs. Alexander Sergienko and Gregg
Jaeger (Boston University), John Myers and Prof. Tai Wu (Harvard),
and Alex Colvin, Chris Lirakis, Oleksiy Pikalo, John Schlafer, Greg
Troxel, and Henry Yeh (BBN). Our interest in QKD networks was
sparked by the prior work of, and discussions with, the quantum
cryptography groups at IBM Almaden and Los Alamos, and by the
kind hospitality of Dr. David Murley several years ago. We also wish
to acknowledge the exemplary contributions of the GAP Optique
group at the University of Geneva, whose papers have been an
inspiration to us.

\nonumsection{References}

\appendix{: \ Analytical Model}
\small
\begin{verbatim}
% File: model.m
%
% Description: Analytic model of QKD throughput
%
% Copyright (c) 2004 by BBN Technologies
%
% This is a Matlab / Octave model of the QKD throughput of a weak-coherent
% source using BB84 through fiber, given various parameters of the system.
% Key parameters include:
%
%       pulseRate       -- the repetition rate of the source, in Hz
%       dutyCycle       -- portion of pulses for payload (vs. header, training)
%       mpn             -- the mean photon number per pulse at Alice
%       fiberLength     -- the length of fiber, in km
%       fiberLoss       -- the attenuation of the fiber, in dB/km
%       rxLoss          -- receiver loss, in dB (Myers eta_rec, as dB)
%       detEff{0,1}     -- detector efficiency, for each detector (eta_det)
%       detLeak{0,1}    -- leakage of other's light into this detector (epsilon)
%       pDark{0,1}      -- probability that detector fires w/ no light
%       pAfter{0,1}     -- probability pulse results in unsuppressed afterpulse
%       residPhase      -- residual phase error, RMS, in radians
%
%       blockSize       -- number of bits in block for EDAC / privacy amplify
%       nEdacSets       -- number of subsets for EDAC
%       estType         -- entropy estimate type ('Bennett', 'Slutsky', 'Myers')
%       confidence      -- probability Eve has more information than estimated
%       siftType        -- type of sifting ('BB84', 'SARG')
%
%       eveChan         -- for PNS, Eve's multiplier on fiberLoss (0=perfect)
%
% This file defines typical values for these variables (which are all
% global variables), and functions which use them to compute the rate
% of detects, errors, and finished bits.  To try different scenarios,
% you can simply modify the global parameters and re-execute the function.

global pulseRate dutyCycle mpn fiberLength fiberLoss rxLoss residPhase
global detEff0 detEff1 detLeak0 detLeak1 pDark0 pDark1 pAfter0 pAfter1
global blockSize nEdacSets estType confidence siftType eveChan

pulseRate = 5e6;                % Alice-Bob link runs at 5MHz
dutyCycle = .8;                 % Measured duty cycle
mpn = .1;                       % Target value (was calibrated recently)
fiberLength = 10.55;            % Length of fiber spool, in km
fiberLoss = .237;               % dB/km for spool, if total = 2.5dB
rxLoss = 10.4;                  % measured loss (dB, average over all paths)
residPhase = 3 * pi/180;        % not measured recently
detEff0 = .117;                 % from analysis of data
detEff1 = detEff0;
detLeak0 = .009;
detLeak1 = detLeak0;
pDark0 = 2.8e-5;
pDark1 = pDark0;
pAfter0 = .001;                 % SW/HW suppression should keep this quite low
pAfter1 = pAfter0;

blockSize = 4096;               % Configured min (average slightly higher)
nEdacSets = 64;                 % Configured
estType = 'Bennett';            % Configured
confidence = 1e-6;              % Hard-wired
siftType = 'BB84';              % Configured

eveChan = 0;                    % Assume Eve has perfect fiber

% sourceRate -- the raw rate of symbols at the source (not counting
% attenuation)

function r = sourceRate
    global pulseRate dutyCycle
    r = pulseRate * dutyCycle;
endfunction

% Utility function to compute the probability of the union of a number of
% independent events

function p = probOr(varargin)
    p = 1;
    for i = (1:nargin)
        p = p * (1-varargin{i});
    endfor
    p = 1 - p;
endfunction

% Here we estimate the probability of the different kinds of detections, and
% turn those probabilities into the sifted rate and QBER.
%
% pmCorr = probability that correct detector fires when bases match
% pmIncorr = probability that incorrect detector fires when bases match
% pwDetect = prob that detector fires when bases wrong (same for both D0 & D1)

function [rate, qber] = siftedRate
    global mpn fiberLength fiberLoss rxLoss residPhase
    global detEff0 detEff1 detLeak0 detLeak1 pDark0 pDark1 pAfter0 pAfter1
    pDark = (pDark0 + pDark1) / 2;
    e = (detEff0*detLeak0 + detEff1*detLeak1) / (detEff0 + detEff1);
    atten = .1^(.1*(fiberLength*fiberLoss + rxLoss));
    c = (detEff0+detEff1)/2 * mpn * atten / (1+detLeak0+detLeak1);
    pwDetect = probOr (pDark, 1-exp(-c*(e + .5)));
    pAfter = pwDetect * (pAfter0 + pAfter1) / 2;
    pwDetect = probOr (pwDetect, pAfter);
    pmCorr = probOr (pDark, pAfter, 1-exp(-c*(e + cos(residPhase/2)^2)));
    pmIncorr = probOr (pDark, pAfter, 1-exp(-c*(e + sin(residPhase/2)^2)));
    pmValid = probOr (pmCorr, pmIncorr);
    rate = pmValid / 2 * sourceRate;
    qber = (pmIncorr - pmCorr*pmIncorr) / pmValid;
endfunction

% EDAC overhead -- this is for the amount of extra information revealed,
% per bit, given the error rate.  This is specifically for the BBN variant of
% Cascade, other protocols are likely to differ slightly.  This also
% represents an average, over many blocks of slightly varying size and
% error rate.  The estimate does not include the error bits themselves.

function ovhd = EDACoverhead (qber)
    global nEdacSets blockSize
    ovhd = qber*(1-log2(qber)) + nEdacSets / blockSize;
endfunction

% entropyEstimate -- this applies the specific entropy estimate chosen
% and then turns it into a fraction of the sifted bits.  The entropy
% estimate here is the information Eve may be assumed to have derived
% from eavesdropping on the single-photon pulses, there is a separate
% function for splitting multi-photon pulses.
%
% It can be tricky to compare estimates because of differing assumptions.
% The entropy derived in Bennett's paper (BBBSS92) refers to the entire
% key string, including error bits -- they are kept in the string and
% accounted for as revealed information during error correction.  The other
% estimates derive entropy on the non-error bits.  In these functions,
% we standardize on Eve's entropy on the non-error bits.
%
% We also explicitly subtract the privacy amplification overhead in the
% estimates, since this is different for the Myers-Pearson estimate (it
% uses Renyi order < 2).

function est = entropyEstimate(qber)
    global estType blockSize confidence
    b = blockSize;
    e = qber*b;
    switch (estType)
    case 'Bennett'
        est = bennett(b,e,confidence);
    case 'Slutsky'
        est = slutsky(b,e,confidence);
    case 'Myers'
        est = myers(b,e,confidence);
    otherwise
        error('Unknown entropy estimate type %s',estType);
    end
    est = est/blockSize;
endfunction

function est = bennett(b,e,confidence)
    t = 2.828427*e;
    dev2 = 6.828427*e;
    conf1 = sqrt(2) * erfinv(1-confidence);
    est = b - e - t - conf1*sqrt(dev2);
    est = est + 2*log2(confidence);
endfunction

function est = slutsky(b,e,confidence)
    conf1 = erfinv(1-confidence);
    eprime = min(e / b + conf1 / sqrt(2*b), 1/3);
    t = (1 - 3*eprime) / (1 - eprime);
    t = (1 + 1.442695*log(1 - 0.5*t*t)) * (b-e);
    dev2 = (b-e)/2;
    est = b - e - t - conf1*sqrt(dev2);
    est = est + 2*log2(confidence);
endfunction

% estimatePNSbits -- how many bits to discard because of "undetectable"
% eavesdropping, i.e. photon-number splitting attacks or unambiguous state
% discrimination (PNS or USD).  This version is essentially Bennett's
% with a more accurate expression for multi-photon pulses.  We assume
% that in all multi-photon pulses, one is captured by Eve and stored until
% the bases are announced.

function mpdisc = estimatePNSbits(sift)
    global mpn detEff0 detEff1 rxLoss
    p0 = exp(-mpn);
    p1 = p0*mpn;
    p2x = 1-p0-p1;
    m = p2x / (p1+p2x);
    mpdisc = m * sift;
endfunction

% estimatePNSgh -- Gilbert & Hamrick's estimate of Eve's information from
% "undetectable" eavesdropping

function mpdisc = estimatePNSgh(sift)
    global fiberLength fiberLoss mpn detEff0 detEff1 rxLoss eveChan
    p0 = exp(-mpn);
    p1 = p0*mpn;
    p2 = p1*mpn/2;
    p2x = 1-p0-p1;
    s2 = sqrt(2);
    y = .1^(.1*(fiberLength*fiberLoss*eveChan + rxLoss)) * (detEff0+detEff1)/2;
    m1 = p2x - 1/(1-y)*(exp(-mpn*y)-exp(-mpn)*(1+mpn*(1-y)));
    m2 = p2*y + 1 - exp(-mpn)*(s2*sinh(mpn/s2)+2*cosh(mpn/s2)-1);
    m3 = p2*y + exp(-mpn)*(sinh(mpn)-s2*sinh(mpn/s2));
    p2k = p2;
    for k = (2:20)
        p2k = p2k * mpn * mpn / (k*(4*k-2));
        m3 = m3 + p2k*max(1-(1-y)^(2*k-1),1-2^(1-k));
    endfor
    m = max([m1,m2,m3]);
    mpdisc = m * sourceRate / 2;
endfunction

% estimatePNSb -- Bennett, et al.'s estimate for Eve's information from
% "undetectable" eavesdropping (BBBSS92)

function mpdisc = estimatePNSb(sift)
    global mpn
    mpdisc = sift*mpn;
endfunction

% distilledRate -- this is the final answer, number of distilled bits per
% second.

function rate = distilledRate
    global confidence
    [sift, qber] = siftedRate;
    ovhd = EDACoverhead(qber);
    ent = entropyEstimate(qber);
    mpd = estimatePNSbits(sift);
    mpd = mpd + sqrt(2)*erfinv(1-confidence) * sqrt(mpd*(1-mpd/sift));
    rate = max(sift*(ent-ovhd) - mpd, 0);
endfunction

% Myers/Pearson entropy estimate
%
% First we find the probability p for which the first k terms of the binomial
% distribution binom(n,i)*p^i*(1-p)^(n-i) sum up to 'confidence', the
% probability that we're wrong.
%
% Then, given this probability, p, the best conditional probability of Eve
% correctly guessing a bit is:
%
%       pe = .5 + sqrt( p/(1-p) * (1 - p/(1-p)) )
%
% Then Eve's least Renyi entropy (order R) for the n-k non-error bits is:
%
%       h(R) = (n-k)/(1-R) * log2(pe^R + (1-pe)^R)
%
% Now from Cachin's paper (Smooth Entropy and Renyi Entropy), theorem 8,
% we know that the amount of smooth entropy (which we can feed into privacy
% amplification) is at least:
%
%       h(R) - log2(m+1) - r/(R-1) - t - 2
%
% where m-log2(m+1) = n+t, and 2^(-r)+2^(-t) = confidence.
%
% If we ignore the negligible effect of t on the value of log(m), the optimal
% values of r and t are:
%
%       r = log2(R/confidence)
%       t = log2(R/((R-1)*confidence))
%
% and the value of m is approximately:
%
%       m = n + t + log2(n+t+1)
% or    m = n + t + log2(n+t+1+log2(n+t+1+log2(n+t+1)))         etc.
%
% In our internal function, we negate this, so we can minimize.

function h = myers_neg_renyi_entropy (r)
    global myers_n myers_k myers_confidence myers_pe
    h = (myers_n - myers_k) / (1-r) * log2(myers_pe^r + (1-myers_pe)^r);
    t = log2(r/((r-1)*myers_confidence));
    h = h - log2(myers_n+t+1+log2(myers_n+t+1+log2(myers_n+t+1)));
    h = h - log2(r/myers_confidence)/(r-1) - t - 2;
    h = -h;
endfunction

% Another internal function -- the sum of the first myers_k terms of the
% binomial distribution, minus myers_confidence (so we can find a zero)

function s = myers_binomtail (p)
    global myers_n myers_k myers_confidence
    k1 = myers_k;
    k2 = myers_n-myers_k;
    if (k1 > k2)
        k1 = k2;
        k2 = myers_k;
    endif

    % Compute the highest term, then go backwards

    if (k1*log(myers_n) < 200)
        % exact if < 10^86
        l = 1;
        for i = 1:k1
            l = l * (myers_n-i+1) / i;
        endfor
        t = l * p^myers_k * (1-p)^(myers_n-myers_k);
    else
        % otherwise use Stirling's approximation
        k1 = k1+1;
        k2 = k2+1;
        n1 = myers_n+1;
        l = 1 - .5*log(2*pi);
        l = l + (1/(n1) - 1/(k1) - 1/(k2)) / 12;
        l = l - (1/(n1)^3 - 1/(k1)^3 - 1/(k2)^3) / 360;
        l = l + (1/(n1)^5 - 1/(k1)^5 - 1/(k2)^5) / 1260;
        l = l + (n1-.5)*log(n1) - (k1-.5)*log(k1) - (k2-.5)*log(k2);
        t = exp(l + myers_k*log(p) + (myers_n-myers_k)*log(1-p));
    endif

    % Now loop back to the beginning, but exit if we stop changing sum

    s = t - myers_confidence;
    for k1 = (myers_k-1:-1:0)
        t = t * (k1+1) * (1-p) / (p * (myers_n-k1));
        s1 = s + t;
        if s1 == s
            break
        endif
        s = s1;
    endfor
endfunction

function entropy = myers(n,k,confidence)
    global myers_n myers_k myers_confidence myers_pe

    % Approximate starting point

    p = 1 - InvBetaApprox(n-k,k,confidence);
    myers_n = n;
    myers_k = k;
    myers_confidence = confidence;

    % Solve for probability p, and compute Eve's probability of guessing

    p = fzero('myers_binomtail',p);
    p = min(p,1/3);
    myers_pe = .5 + sqrt( p/(1-p) * (1 - p/(1-p)) );

    % Maximize entropy measure over Renyi order R

    r = fminbnd('myers_neg_renyi_entropy',1.01,2);

    % Return the maximized entropy

    entropy = -myers_neg_renyi_entropy(r);
endfunction

% Abramowitz and Stegun approximation for the inverse of the incomplete
% Beta function

function v = InvBetaApprox(a,b,p)
    y = sqrt(2) * erfinv(1-2*p);
    l = y*y/6 - .5;
    a1 = 1/(2*a-1);
    b1 = 1/(2*b-1);
    h = 2/(a1+b1);
    w = y*sqrt(h+l)/h - (b1-a1)*(l+5/6-2/(3*h));
    v = a/(a+b*exp(2*w));
endfunction
\end{verbatim}


\begin{thebibliography}{000}
\bibitem{Gisin02}
N. Gisin, Ribordy, G, W. Tittel, and H. Zbinden (2002), ``Quantum
cryptography,'' {\it Reviews of Modern Physics}, {\bf 74}.
\bibitem{SSPD01}
G. N. Gol'tsman, O. Okunev, G. Chulkova, A. Lipatov,
A. Semenov, K. Smirnov, B. Voronov, A. Dzardanov,
C. Williams, and R. Sobolewski (2001), ``Picosecond superconducting
single-photon optical detector,'' {\it Applied Physics Letters},
{\bf 79}(6), pp. 705--707.
\bibitem{NIST02}
C. J. Williams, X. Tang, M. Heikkero, J. Rouzaud, R. Lu,
A. Goedecke, A. Migdall, A. Mink, A. Nakassis,  L. Pibida,
J. Wen, E. Hagley and C. W. Clark (2002), ``A High Speed Quantum
Communications Testbed,'' Proc. SPIE International Symposium
of Optical Science and Technology, July 2002.
\bibitem{Myers04}
J. M. Myers, T. Wu and D. Pearson (2004),
``Entropy estimates for individual attacks on the BB84 protocol
for quantum key distribution,''
Proceedings of SPIE Defense and Security 2004,
vol. 5105---Quantum Information and Computation, April 2004.
\bibitem{BBBSS92}
C. Bennett, F. Bessette, G. Brassard, L. Salvail and J. Smolin (1992),
``Experimental quantum cryptography,'' {\it J. Cryptology}, {\bf 5}(1),
pp. 3--28.
\bibitem{GH00}
G. Gilbert and M. Hamrick (2000), ``Practical Quantum Cryptography: A
Comprehensive Analysis (Part One),'' quant-ph/0009027.
\bibitem{LANL00}
W. T. Buttler, R. J. Hughes, S. K. Lamoureaux, G. L. Morgan,
J. E. Nordholt and C. G. Peterson (2000), ``Daylight quantum key
distribution over 1.6 km,'' {\it Phys. Rev. Lett.} {\bf 84}, pp.
5652--5655, quant-ph/0001088.
\bibitem{Bethune04}
D. Bethune and W. Risk (2004), ``Autocompensating quantum
cryptography,'' {\it New Journal of Physics}, {\bf 4}, pp. 42.1--42.15,
quant-ph/0204144
\bibitem{Elliott02}
C. Elliott (2002), ``Building the quantum network,'' {\it New Journal
of Physics}, {\bf 4}, pp. 46.1--46.12.
\bibitem{BBN03}
C. Elliott, D. Pearson and G. Troxel (2003), ``Quantum Cryptography in
Practice,'' Proc. ACM SIGCOMM 2003.
\bibitem{LANL96}
Hughes, R. et al (1996), ``Quantum cryptography over underground
optical fibers,'' {\it in N. Koblitz}, editor, Advances in Cryptology---CRYPTO
'96, of Lecture Notes in Computer
Science, {\bf 1109} pp. 329--342.
\bibitem{SARG03}
V. Scarani, A. Acin, G. Ribordy and N. Gisin (2003), ``Quantum
cryptography protocols robust against photon number splitting
attacks,'' ERATO Conference on Quantum Information Science,
September 4--6, 2003, Niijima-kaikan, Kyoto, Japan.
\bibitem{Slut98}
B. Slutsky, R. Rao, P. Sun, L. Tancevski and S. Fainman (1998),
``Defense frontier analysis of quantum cryptographic systems,''
{\it Applied Optics}, {\bf 37}, no. 14, pp. 2869--2878.
\bibitem{BLMS00}
G. Brassard, N. L\"{u}tkenhaus, T. Mor and B. C. Sanders (2000),
``Security Aspects of Practical Quantum Cryptography,'' {\it Phys. Rev.
Lett.} {\bf 85}, pp. 1330--1333, quant-ph/9911054.
\bibitem{LutPRA}
J. Calsamiglia, S. M. Barnett and N. L\"{u}tkenhaus (2002), ``Conditional
beam-splitting attack on quantum key distribution,'' {\it Phys. Rev. A}
{\bf 65}, quant-ph/0107148.
\bibitem{Curty}
M. Curty, and N. L\"{u}tkenhaus, ``Practical quantum key distribution:
On the security evaluation with inefficient single-photon detectors,''
quant-ph/0311066.
\bibitem{Lut00}
N. L\"{u}tkenhaus (2000), ``Security against individual attacks for realistic
quantum key distribution,'' {\it Phys. Rev. A} {\bf 61}, quant-ph/9910093.
\end{thebibliography}
\end{document}